\documentclass[aps, prl, reprint, superscriptaddress]{revtex4-2} 
\usepackage[utf8]{inputenc}
\usepackage[sort&compress]{natbib}
\usepackage{ulem}
\usepackage{overpic}
\usepackage{bm}
\usepackage{times}
\usepackage{amssymb,amsbsy,amsmath,amsfonts}
\usepackage{graphicx}
\usepackage{color}
\usepackage{booktabs}
\usepackage{makecell}

\usepackage{rotating}
\usepackage{srcltx}
\usepackage{slashed}
\usepackage{subfigure}
\usepackage{multirow}
\usepackage{verbatim}
\usepackage{hyperref}
\usepackage{tabularx}

\DeclareUnicodeCharacter{3002}{HEREHEREHERE}

\newcolumntype{C}{>{\centering\arraybackslash}X}

\begin{document}

\title{
Nanosecond-Scale Proton Emission from Triaxially Deformed \texorpdfstring{$^{148}$Lu}{148Lu} Predicted with High Accuracy \texorpdfstring{$Q_{\rm p}$}{Qp} Value via Novel Bayesian Evaluation}

\author{Lin-Xing Zeng}
\affiliation{School of Physics, Beihang University, Beijing 102206, China}

\author{Qi Lu}
\affiliation{School of Physics, Beihang University, Beijing 102206, China}

\author{Kaiyuan Zhang}\email{zhangky@caep.cn}
\affiliation{National Key Laboratory of Neutron Science and Technology, Institute of Nuclear Physics and Chemistry, China Academy of Engineering Physics, Mianyang, Sichuan 621900, China}

\author{Shi-Sheng Zhang}\email{zss76@buaa.edu.cn}
\affiliation{School of Physics, Beihang University, Beijing 102206, China}

\date{\today}

\begin{abstract}
The half-life of the odd-odd deformed proton emitter $^{148}$Lu is predicted to be $196_{-129}^{+420}$ ns via the Wentzel-Kramers-Brillouin (WKB) approximation,
in which the potential is extracted from the triaxial relativistic Hartree-Bogoliubov theory in continuum (TRHBc) 
and the proton decay energy $Q_{\rm p}$ is computed as 2.015(89) MeV by the Bayesian Neural Network - Beihang (BNN-BH) model for the first time.
As a decisive factor, the uncertainty of $S_{\rm p}$ has been improved from 411 keV (Bayesian Machine Learning, BML) to 89 keV (BNN-BH) by taking the ensemble uncertainty into account and confining the error estimation to the neighboring nuclei.
In consequence, the magnitude of the half-life's uncertainty can be reduced from 4 orders to 1 order, compared to that ($5.5_{-5.3}^{+636}$ ns) with $S_{\rm p}$ from the BML model.
We also found that the range of half-life predicted by the TRHBc + WKB approach is consistent with those from the deformed relativistic Hartree-Bogoliubov theory in continuum (DRHBc) + WKB approach, and with those from an empirical formalism with the $S_{\rm p}$ obtained with the BNN-BH model.
Furthermore, the means from the above 3 ways agree well with the experimental data for $^{149}$Lu, which gives us confidence 
to recommend a measurement of the half-life of proton emitter $^{148}$Lu.

\end{abstract}


\maketitle



{\it Introduction.---}The origin of elements heavier than iron relates to helium burning in Red Giant stars ({\it s}-process) and supernova explosions ({\it r}- and {\it p}- processes)
~\cite{meyer_r-_1994, Arnould:2001xs, kappeler_stellar_2012, kappeler_reaction_2011}.
Half-lives of the proton-rich nuclei~\cite{Wang_2021, Yuan:2023myz} during the {\it p}-process based on the {\it s}-nuclides seeds close to the drip line ({\it i.e.} {\it p}-nuclei) play an important role in nucleosynthesis by changing the nuclear reaction pathways in massive stars~\cite{zhang_new_2024}.
Meanwhile, the proton emission beyond the drip line aroused great attention since the measurement of the shortest ground state proton emitter's half-life, $T_{1/2} = 450^{+170}_{-100}$ ns for $^{149}$Lu~\cite{PhysRevLett.128.112501, noauthor_squashed_2022}.
Therefore, the search for a proton emitter candidate with an even shorter but measurable half-life -- perhaps $^{148}$Lu~\cite{Xiao:2023uld, lu_triaxial_2024} -- is desired, which is sensitive to the structure information~\cite{DELION2006113,Ferreira:2001zz}, including the deformation, the orbital angular momentum of the emitted proton,
and decay energy.

There are several methods used to calculate the half-lives of nucleon emitters.
A direct and effective way is to solve the multiple-coupled-channel-equations to obtain the radial components of all allowed decay channels in the well-deformed nuclei, 
which can be used to obtain the partial widths of those channels~\cite{delion_exact_2015, delion_systematics_2006, peltonen_-decay_2008, ni_microscopic_2009, ni_new_2010, zhang__2011, delion_coupled_2018}.
However, the experimental excitation energies of the daughter nuclei are required by the matrix element of the equations, which are not available for $^{148}$Lu.
Alternatively, one can combine the density functional theory (DFT)~\cite{meng_relativistic_2016, bender_self-consistent_2003}
with the Wentzel-Kramers-Brillouin (WKB) approximation to calculate the half-lives of proton emitters. 
This approach works well for $^{149}$Lu~\cite{Xiao:2023uld, lu_triaxial_2024}, where nuclear masses are available for neighboring nuclei;
the half-life is determined by the exponential term with the potential from DFT and the decay energy $Q_{\rm p}$ from the nuclear mass difference.
Hundreds of keV deviation from $Q_{\rm p}$ will lead to orders of magnitude difference in proton emitter's half-life.
Therefore, the prediction of proton separation energy ($S_{\rm p}=-Q_{\rm p}$)~\cite{kokkonen_new_2025} with high accuracy is strongly desired when the relevant nuclear masses are not available.





In the last decade, numerous theoretical models have been adopted to predict nuclear masses.
Within the framework of nonrelativistic DFT, 
the root-mean-square deviation (RMSD) of calculated masses from the data can reach 561 keV for 2353 nuclei, with typically more than 20 model parameters~\cite{PhysRevC.93.034337}.
On the relativistic side, an accuracy of about 1400 keV for nuclear masses has been achieved by the covariant DFT with approximately 10 model parameters~\cite{Zhang2022ADNDT,Guo2024ADNDT}
Among the phenomenological models, the finite-range droplet model (FRDM)~\cite{Mller2012NewFD} with more than 30 model parameters yields an accuracy of 570 keV for the masses complied in the atomic mass evaluation AME2012~\cite {wang_ame2012_2012}.
So far, the predictions of nuclear masses with the lowest stated RMSD ($<300$ keV) are obtained with the 20-free-parameter Weizs$\ddot{\rm a}$cker-Skyrme 4 (WS4) model~\cite{wang_surface_2014}.
However, to constrain the half-lives of the proton emitters within 2 orders of magnitude, the accuracy for nuclear masses or separation energies is required to be less than 100 keV.

Recently, Machine Learning (ML) methods have been widely applied to nuclear physics due to their powerful ability of learning and prediction.
Among the models for mass predictions, 
Kernel Ridge Regression (KRR), which performs well on the small dataset and extrapolation, has reduced the RMSD to 128 keV~\cite{WU2021136387}.
As a universal approximator~\cite{HORNIK1989359}, Artificial Neural Network (ANN) is improved to the accuracy of 260 keV~\cite{zeng_nuclear_2024} via the constraint of the Garvey-Kelson relations~\cite{garvey_set_1969}.
A Kolmogorov-Arnold network (KAN) is used to achieve an RMSD of 260 keV by deriving symbolic expressions from the mass data~\cite{PhysRevC.111.024316}.
Up to now, Bayesian Machine Learning (BML) model~\cite{niu_nuclear_2022} produces the lowest quoted RMSD of ML approaches in the predictions of masses with the uncertainty.

These models commonly have a trade-off between the accuracy and overfitting.
Even for the best phenomenological WS4 model, which was fit on data from AME2012, the RMSD
grows from 468 keV (AME2016~\cite{CPC:10.1088/1674-1137/41/3/030003}) to 1295 keV (with extra 120 nuclei in AME2020~\cite{wang_ame_2021}),  indicating limited capability of extrapolation.
Therefore, it is necessary to balance high accuracy and overfitting to enhance the extrapolation ability in the predictions of $S_{\rm p}$, particularly for the exotic nuclei around the drip lines. 
Based on this idea, we propose a novel Bayesian Neural Network-Beihang (BNN-BH) model which can give predictions with high accuracy and mitigate the overfitting problem.

In addition to $S_{\rm p}$ predicted by our BNN-BH model, 
microscopic nuclear potentials $V(r,\theta,\phi)$ used to determine the half-life are extracted from the deformed relativistic Hartree-Bogoliubov theory in continuum (DRHBc) theory~\cite{Zhou:2009sp, Li:2012gv, Li:2012xaa, DRHBcMassTable:2020cfw, DRHBcMassTable:2022rvn}.
This approach considers axial deformation, pairing correlations, blocking, and continuum effects in a self-consistent way, which has achieved great success in studying exotic nuclear phenomena near and beyond the drip line~\cite{Meng2015JPG, Sun2018PLB, Zhang2019PRC, Sun2020NPA, Yang2021PRL, Zhang2021PRC(L), Pan2021PRC, Sun2021PRC(1), Zhang2023PLB, Zhang2023PRC(L1), Pan2024PLB, Zhang2025PLB, Zhang2025AAPPS}.
In particular, a unified description from structure to reaction dynamics of one-neutron halo nuclei, such as $^{31}$Ne and $^{37}$Mg, has been successfully realized with the DRHBc theory combined with the Glauber model~\cite{zhong2021studydeformedhalonucleus, AN2024138422, Wang2024EPJA, Zhang:2021fpl,Zhong:2021yhm}.
As an extension of the DRHBc theory, the triaxial relativistic Hartree–Bogoliubov theory in continuum (TRHBc) theory~\cite{Zhang2023PRC(L2), gk2w-cblb}, has recently been developed and applied to examine the effects of triaxial deformation on the proton emission of $^{149}$Lu~\cite{lu_triaxial_2024}.
There are some other calculations for most lutetium isotopes and their neighbours in the $A\approx{150}$ mass region are triaxial~\cite{Chen2019PRC}.


In this Letter, we aim at confining the half-life of 
proton emitter $^{148}$Lu to high accuracy for future measurement. 
We first review the TRHBc theory which determines the shape of the candidate and provides the potential for WKB calculations.
Then, we introduce our BNN-BH model and demonstrate that it can predict $S_{\rm p}$ of $^{148}$Lu within the uncertainty of 90 keV.
We subsequently analyze the potential energy surface, and compare $S_{\rm p}$ from the BNN-BH model to that from the BML model with different uncertainties, and predict the half-life of $^{148}$Lu. 
Finally, we make a brief summary.






{\it The TRHBc + WKB theory.}---{The TRHBc theory provides a powerful tool to describe exotic nuclei nearby the drip lines, {\it e.g.}, examining the possibility of a proton halo in the drip line nucleus $^{22}$Al~\cite{Zhang2024PRC, yt6n-bp9f}, and revealing the triaxial shape of the one-proton emitter $^{149}$Lu~\cite{lu_triaxial_2024}. The theoretical framework of TRHBc has been outlined in Refs.~\cite{Zhang2023PRC(L2), lu_triaxial_2024}, with a comprehensive formalism detailed in Ref.~\cite{gk2w-cblb}. Below, we summarize several key aspects essential to the subsequent discussion.

In the TRHBc theory, the RHB equations for quasiparticles are solved using a spherical Dirac Woods-Saxon basis~\cite{Zhou:2003jv, Zhang2022PRC}. This basis exhibits an appropriate asymptotic behavior at large distances, enabling an adequate description of the possible large spatial extension in exotic nuclei. Moreover, each basis state--obtained by solving the Dirac equation with a spherical Woods-Saxon potential--carries good quantum numbers $nljm$, facilitating the analysis of the main components of single-particle orbitals. These orbitals are eigenstates of the density matrix, with their occupation probabilities given by the corresponding eigenvalues~\cite{Peter1980Book}. 
The relativistic mean field for neutrons or protons is constructed as a sum of vector and scalar potentials,
which are expanded in terms of spherical harmonics and self-consistently determined from various nuclear densities~\cite{Xiang2023Symmetry}. The RHB equations are solved iteratively, with quasiparticle wave functions, densities, and potentials updated at each step until full self-consistency is achieved. Upon convergence, physical quantities including binding energies $E_{\rm B}$, root-mean-square radii, nuclear densities, mean-field potentials, and single-particle orbitals can be extracted.}

Inspired by our previous successful studies of the neutron decay widths~\cite{2022EPJWC.26011037X}, we adopt the simple and effective semi-classical WKB approximation to calculate the proton emission half-life, {\it{i.e.}},
\begin{align}\label{eq:halflife}
    T_{1/2} = \frac{\hbar \ln{2}}{S_F \it{\Gamma}}, 
\end{align}
where the spectroscopic factor $S_F$ for a nucleus with odd atomic number $Z$ in the independent-quasiparticle approximation can be written as~\cite{Sorensen:1966zz}$S_F =u^2$,
in which $u^2$ denotes the unoccupied probability of the corresponding orbital in the daughter nucleus and can be obtained self-consistently from the TRHBc framework. 
The final decay width $\it{\Gamma}$ is obtained by averaging $\it{\Gamma}(\theta, \phi)$ in all directions. 

In the WKB approximation, $\it{\Gamma}(\theta, \phi)$ is approximated  as
\begin{align}\label{eq:width}
    {\it{\Gamma}}(\theta,\phi) = N \frac{\hbar^2}{4\mu}\exp{\left[-2\int_{r_2}^{r_3}k(r,\theta, \phi) \text{d} r\right]},
\end{align}
where the momentum $k(r,\theta,\phi)=\sqrt{2\mu|Q_{\rm p} - V(r,\theta,\phi)|}$, $\mu$ refers to the reduced mass of the emitted proton and the daughter nucleus, $r$ denotes the distance between the proton and the mass center of the daughter nucleus, and $r_i$ corresponds to the $i$-th turning point at barrier $V(r,\theta,\phi)$ with respect to the decay energy $Q_{\rm p}$. The quasiclassic bound-state normalization factor $N$ in Eq.(\ref{eq:width}) is defined by~\cite{Buck:1992zza}
\begin{align}\label{eq:nor}
    N^{-1}(\theta,\phi) = \int_{r_1}^{r_2} \frac{\text{d}r}{k(r,\theta,\phi)} \cos^2 \left[ \int_{r_1}^r \text{d} r' k(r',\theta,\phi)  -\frac{\pi}{4} \right].
\end{align}
The potential barrier reads
\begin{align}\label{eq:barrier}
    V(r,\theta,\phi) = V_N(r,\theta,\phi) + V_C(r,\theta,\phi) + \frac{\hbar^2}{2\mu} \frac{l(l+1)}{r^2},
\end{align}
where $V_{N/C}(r,\theta,\phi)$ refers to the nuclear/Coulomb potential and can be self-consistently calculated by the 
TRHBc theory.

{\it The BNN-BH model.}---Recently, BNN models~\cite{neal_bayesian_1996} have been widely used in nuclear physics; these map the input vector $\mathbf{x}$ to the prediction vector $\mbox{BNN}(\mathbf{x})$ through several hidden layers. 
Here, our fully-connected BNN model contains 2 hidden layers both with 32 neurons.
The input vector $\mathbf{x}$ propagates from the input layer to the output layer according to $\mathbf{x}_{i} = f(\mathbf{w}_{i}\mathbf{x}_{i-1} + \mathbf{b}_{i})$, in which $f$ is a pointwise nonlinear activation function, the weight matrix $\mathbf{w}_{i}$ and the bias vector $\mathbf{b}_{i}$ are the learnable parameters $\mathbf{\Theta}$ of the {\it i}-th layer, and $i$ runs over the hidden layers and the output layer.
The BNN model is trained to minimize the loss function $\mathcal{L}$
by updating $\mathbf{\Theta}$ (through the backpropagation algorithm~\cite{rumelhart_learning_1986}) as follows,
\begin{equation}\label{eq:loss min}
  \mathbf{\Theta}^{*} = \underset{\mathbf{\Theta}}{\arg \min} \mathcal{L}\left( \mbox{BNN}(\mathbf{x};\mathbf{\Theta}), \mathbf{y} \right) ,
\end{equation}
where $\mathbf{y}$ denotes the label (true value) corresponding to the input $\mathbf{x}$ in the training set.
The parameters $\mathbf{\Theta}$ are described as posterior distribution $p(\mathbf{\Theta}|D)$ learned from the dataset $D = \{ (\mathbf{x}^{i}, \mathbf{y}^{i}) \}$ in the BNN model,
so $\mbox{BNN}(\mathbf{x};\mathbf{\Theta})$ is also described probabilistically.
Here, $p(\mathbf{\Theta}|D)$ is obtained via the Bayes theorem
$p(\mathbf{\Theta}|D) = \frac{p(D|\mathbf{\Theta})p(\mathbf{\Theta})}{p(D)}$,
where $p(D|\mathbf{\Theta})$ is the likelihood of $D$ under $\mathbf{\Theta}$,
$p(\mathbf{\Theta})$ is the prior distribution of $\mathbf{\Theta}$ adopting Gaussian function $\mathcal{N}(0, 0.1^{2})$,
and $p(D) = \int d\mathbf{\Theta}p(D|\mathbf{\Theta})p(\mathbf{\Theta})$ is the evidence for normalization.
In practice, $p(D)$ is approximated by minimizing the Kullback-Leibler ($KL$) Divergence $KL[q(\mathbf{\Theta})||p(\mathbf{\Theta})] = \int d\mathbf{\Theta} q(\mathbf{\Theta}) \log q(\mathbf{\Theta}) / p(\mathbf{\Theta})$,
where $q(\mathbf{\Theta}) = \mathcal{N}(\mu_{\mathbf{\Theta}}, \sigma_{\mathbf{\Theta}}^{2})$ is the parameterization of $p(\mathbf{\Theta}|D)$, based on Variational Inference~\cite{blundell2015weightuncertaintyneuralnetworks}.
To fit the means of the predictions, we also   minimize Mean Squared Error $MSE = \frac{1}{N} \sum\limits_{i=1}^{N} \left(\frac{\overline{\mbox{BNN}(\mathbf{x}^{i};\mathbf{\Theta})} - \overline{\mathbf{y}^{i}}}{\sigma_{\mathbf{y}^{i}}}\right)^{2}$, where $\sigma_{\mathbf{y}}$ is the uncertainty of $\mathbf{y}$.
Therefore, the loss function in Eq.~(\ref{eq:loss min}) is defined as $\mathcal{L} = MSE + 0.01 KL$.

We choose 7 features for the input vector $\mathbf{x}$:
$Z$ (proton number), $N$ (neutron number), $Z_{\rm EO}$, $N_{\rm EO}$ (set as 1 or 0 when $Z$ ($N$) is odd or even, representing the odd-even staggering), $\Delta Z$, $\Delta N$ (differences between $Z$ ($N$) and the nearest magic number, representing shell effect) and $E_{\rm sym}$ (symmetry energy term in WS4 model~\cite{wang_surface_2014}).
The residuals of $S_{\rm p}$ and $E_{\rm B}$ between the theoretical predictions and the experimental data are regarded as the label values $\mathbf{y}$.
Here, the theoretical models include WS4, Bethe-Weizs$\ddot{\rm a}$cker (BW)~\cite{KIRSON200829}, Finite-Range Droplet Model 2012 (FRDM2012)~\cite{Mller2012NewFD}, Koura-Tachibana-Uno-Yamada 05 (KTUY05)~\cite{koura_nuclidic_2005} and Relativistic Mean Field (RMF)~\cite{10.1143/PTP.113.785}.
Experimental data are taken from National Nuclear Data Center (NNDC)~\cite{noauthor_nndc_nodate} ($S_{\rm p}$) and AME2020 ($E_{\rm B}$).
Based on those above, there are 3056 nuclei in the dataset for training set (80\%) and test set (20\%).
The mean $\overline{O}$ and uncertainty $\sigma$ of the object $O$ ({\it e.g.} $S_{\rm p}$ and $E_{\rm B}$) are computed as below,
\begin{equation}\label{eq:mean sigma BNN}
  \overline{O}(Z,N) = \frac{\sum\limits_{i}^{n_{\rm model}}\frac{\overline{O}_{i}(Z,N)}{\sigma_{i}^{2}(Z,N)}}{\sum\limits_{i}^{n_{\rm model}}\frac{1}{\sigma_{i}^{2}(Z,N)}} ,
  \sigma^{2}(Z,N) = \frac{n_{\rm model}}{\sum\limits_{i}^{n_{\rm model}}\frac{1}{\sigma_{i}^{2}(Z,N)}} ,
\end{equation}
where $\overline{O}_{i}$ and $\sigma_{i}$ refer to the mean and the standard deviation of the residual by the {\it i}-th model, respectively.
The choice of the weight $1 / \sigma_{i}^{2}$ exhibits the importance of the predictions with smaller uncertainties.

We further consider the uncertainty $\sigma_{\rm en}$ based on the ensemble learning of the BNN models.
Since some sources of error are unknown, we add the factor $\chi_{\nu}$ into the total uncertainty $\sigma_{total} = \chi_{\nu} \sigma'$,
in which $\chi_{\nu}^{2} = \sum\limits_{i=1}^{N}\frac{1}{N}\frac{\left(O_{i}-L_{i}\right)^2}{\sigma'^{2}_{\rm \it{i}}}$ only iterates the lutetium isotopes ($O_{i}$ refers to $S_{\rm p}$ of the {\it i}-th nuclei, $L_{i}$ represents the experimental data),
and the known uncertainty is defined as $\sigma'^{2} = \sigma_{\rm en}^{2} + \sigma^{2}$.
Concretely, $\sigma_{\rm en}$ is written by 
\begin{equation}\label{eq:ensemble uncertainty}
  \sigma_{\rm en}^{2} = \frac{ \sum\limits_{\rm seed} \left( S_{\rm p}^{\rm seed}-S_{\rm p}^{\rm BNN-BH} \right)^{2} / \sigma_{\rm seed}^{2} } { \sum\limits_{\rm seed} \left. 1 \middle/ \sigma_{\rm seed}^{2} \right. }.
\end{equation}
We construct 20 BNN models initialized by different random seeds.
These 20 means $S_{\rm p}^{\rm seed}$ and uncertainties $\sigma_{\rm seed}$ are combined to yield the mean $S_{\rm p}^{\rm BNN-BH}$ and the uncertainty $\sigma$ with the same procedure of Eq.~(\ref{eq:mean sigma BNN}).
Compared to the available description, {\it e.g.} the BML model, we emphasize the contribution from $\sigma_{\rm en}$ in the uncertainty, and we especially confine $\chi_{\nu}$ to the neighborhood nuclei.

In practice, we evaluate the BNN-BH model by two factors, {\it i.e.} the accuracy via RMSD on measured lutetium  isotopes and an estimate of the size of the overfitting problem.
The RMSD is defined as $\sigma_{\rm rms} = \sqrt{ \sum\limits_{i=1}^{N}\frac{\left( O_{i}-L_{i} \right)^{2}}{N}}$,
and the magnitude of overfitting is estimated by $r_{\rm set}=\sigma_{\rm rms}^{\rm test}/\sigma_{\rm rms}^{\rm train}$.
Here, $\sigma_{\rm rms}^{\rm train}$ and $\sigma_{\rm rms}^{\rm test}$ are the RMSDs of the training set and the test set, respectively.
We suppose the case of $r_{\rm set} > 1.2$ as severe overfitting problem.













\begin{figure}[htbp]
    \centering
    \includegraphics[width=8cm]{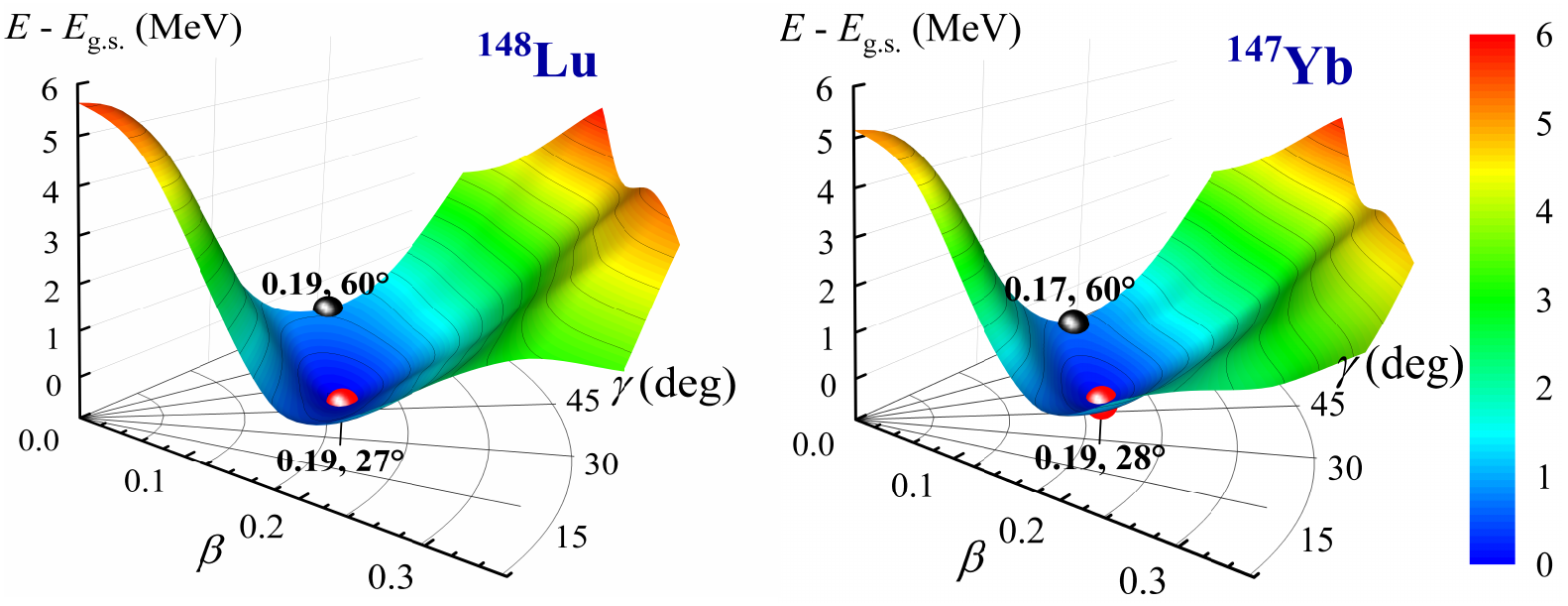}
    \caption{Potential energy surfaces for $^{148}$Lu and $^{147}$Yb in the $\beta$-$\gamma$ plane from the TRHBc theory. The energy interval between each contour line is 0.4 MeV. All energies are normalized with respect to the energy of absolute minimum (triaxial ground state) indicated by a red ball. The ground state deformation predicted by the DRHBc theory is denoted by a black ball.}
    \label{pic:surface energy}
\end{figure}

{\it Results and discussion.}---
{Figure~\ref{pic:surface energy} displays the potential energy surfaces (PESs) for the proton emitter $^{148}$Lu and its daughter nucleus $^{147}$Yb, obtained from the deformation constrained by TRHBc calculations. Analogous to $^{149}$Lu and several other proton-rich lutetium isotopes~\cite{Chen2019PRC}, the ground state of $^{148}$Lu is characterized as a triaxial shape with deformation parameters ($\beta,\gamma$) = ($0.19,27^\circ$). Due to the removal of only one proton, the PES and triaxial shape of $^{147}$Yb are very similar to those of $^{148}$Lu. In the absence of triaxial degrees of freedom, the DRHBc theory predicts oblate deformations for their ground states; however, the PESs clearly show that these oblate configurations correspond to saddle points rather than true minima. The energy difference between the triaxial ground state and the oblate saddle point exceeds 0.8 MeV, underscoring the essential role of triaxiality in the microscopic description of $^{148}$Lu and $^{147}$Yb. In particular, the significant triaxial deformation ($\gamma \approx 30^\circ$) shapes the nuclear potential and tunneling barrier, which may have a non-negligible impact on the decay width of  proton emitter~\cite{lu_triaxial_2024}.
}



\begin{table}[htbp]
    \centering
    \caption{RMSDs (keV) of $S_{\rm p}$ for training set, test set and lutetium isotopes with the BNN-BH and BML model, together with $r_{\rm set}$.}\label{table:BNN results}
    \begin{tabularx}{\linewidth}{XXXXX}
      \hline\hline
        Models & Training set  & Test set & Lutetium isotopes & $r_{\text{set}}$ \\ \hline
        BNN-BH & 156 & 182 & 74 & 1.17 \\ 
        BML~\cite{niu_nuclear_2022} & 220 & 200 & 128 & - \\
      \hline\hline
    \end{tabularx}
\end{table}

As the best model to predict the nuclear masses with the uncertainty,
the predictions of the BML model~\cite{niu_nuclear_2022} are used to calculate the RMSDs of $S_{\rm p}$ for our training set, test set and lutetium isotopes.
Those results are compared with ours, and listed in Table~\ref{table:BNN results}.
Note the BML model obtains the RMSD of 84 keV only for the measured data from AME2016 with errors less than 100 keV,
while our BNN-BN model adopts all the available $S_{\rm p}$ from NNDC.
Hence, the RMSDs of the BML model in our training and test sets are larger than 84 keV.
It can be seen that the RMSDs of $S_{\rm p}$ from the BNN-BH model are smaller than those from the BML model.
Furthermore, the calculated $r_{\rm set}$ from our model is less than 1.2,
indicating that the magnitude of overfitting is acceptable.
Based on the higher accuracy and reasonable overfitting, our model can be safely applied to the nuclei with unknown $S_{\rm p}$, such as $^{148}$Lu.

\begin{table*}[htbp]
\centering
\caption{
Half-life of the proton emitter $^{148}$Lu. Note that for theories the orbital refers to the main component of the single-proton level occupied by the valence proton.
The $Q_{\rm p}$-values used in our half-life calculations are taken from the BNN-BH and BML models, which give $Q^{\mathrm{BNN-BH}} = 2.015(89)$ MeV and $Q^{\mathrm{BML}} = 2.745(411)$ MeV.
}
\label{tb:result}
\begin{tabularx}{\linewidth}{CCCCC}
\hline
\hline
\multirow{2}*{$^{148}$Lu} & \multicolumn{2}{c}{DRHBc} & \multicolumn{2}{c}{TRHBc} \\
\cline{2-5}
{} & BNN-BH & BML & BNN-BH & BML \\
\hline
Orbital  & \multicolumn{2}{c}{$1h_{11/2}$} & \multicolumn{2}{c}{$1h_{11/2}$} \\

Deformation & \multicolumn{2}{c}{$\beta_2 = -0.19 $} & \multicolumn{2}{c}{$(\beta_2,\gamma)=(0.19,27^\circ)$} \\
$S_F$ & \multicolumn{2}{c}{0.77} & \multicolumn{2}{c}{0.55} \\
lg$\mathit{\Gamma}$ & $-14.36_{-0.47}^{+0.50}$ & $-12.81_{-2.07}^{+1.59}$ & $-14.37_{-0.47}^{+0.50}$ & $-12.82_{-2.07}^{+1.59}$ \\
$T_{1/2}$ (ns) & $137_{-90}^{+294}$ & $3.8_{-3.7}^{+445}$ & $196_{-129}^{+420}$ & $5.5_{-5.3}^{+636}$ \\
\hline\hline
\end{tabularx}
\end{table*}

\begin{figure}[htbp]
    \centering
    \includegraphics[width=8cm]{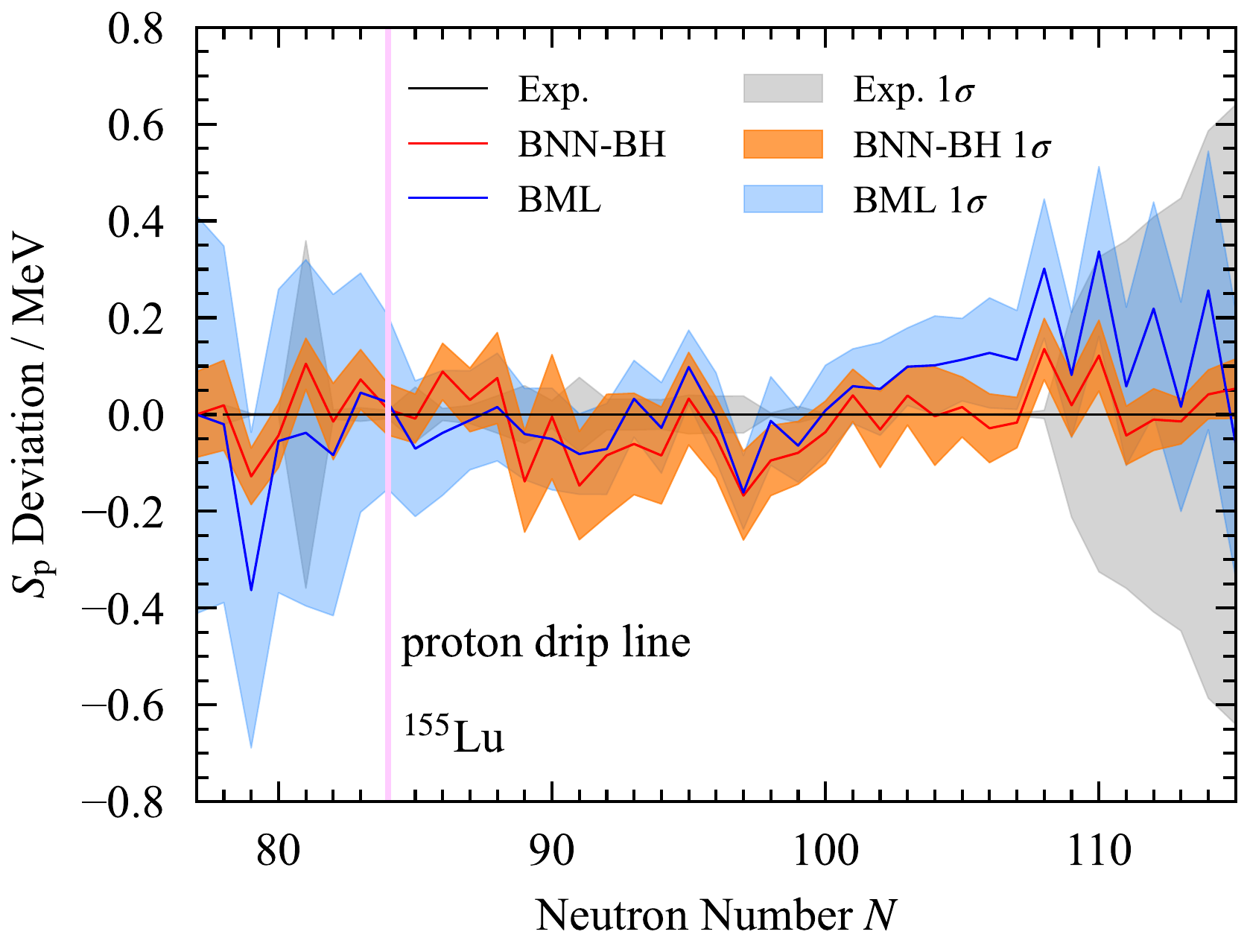}
    \caption{
    Deviations of $S_{\rm p}$ in $^{148-186}$Lu isotopes with the BNN-BH and the BML~\cite{niu_nuclear_2022} models, in comparison with the available data.
    The solid red (blue) lines with uncertainties show the deviations of the predictions from the BNN-BH (BML) models.
    Shaded regions represent the $1\sigma$ uncertainties.
    The solid pink line corresponds to the proton drip line ($^{155}$Lu).
    Experimental data are taken from Ref.~\cite{PhysRevLett.128.112501} ($^{149}$Lu) and NNDC~\cite{noauthor_nndc_nodate} ($^{150-186}$Lu).
    }
    \label{pic:Lu-chain dev}
\end{figure}

To completely display the ability of these two models to predict $S_{\rm p}$ in lutetium isotopes,
we plot the deviations of $S_{\rm p}$ in Fig.~\ref{pic:Lu-chain dev}.
Note the experimental $S_{\rm p}$ of $^{148}$Lu is absent.
It can be seen that the variation of the predicted distributions by these two models are quite similar for $^{160-172}$Lu isotopes in the middle of mass region.
Whereas, the uncertainties of the BML model are significantly enlarged in other regions, especially for $^{148-154}$Lu beyond the proton drip line.
The uncertainties of these lutetium isotopes are at least 3.9 times of our results in orange band, which are below 130 keV for the whole region and within 90 keV for $^{148}$Lu.
It is worth noting the $\sigma_{\rm en}$ of $^{148}$Lu by our model is 55 keV,
while the BML model does not consider such an important source of error.
As a result, $\chi_{\nu}$ is accordingly reduced from 2.1 (BML) to 1.58 (BNN-BH), which means that the uncertainty of our model is more reliable due to the decrease of unknown error sources.
Therefore, we naturally extrapolate to the lutetium isotopes far from the stability line to predict the half-life of proton emitters with effectively constrained values of $S_{\rm p}$.



\begin{figure}[htbp]
    \centering
    \includegraphics[width=8cm]{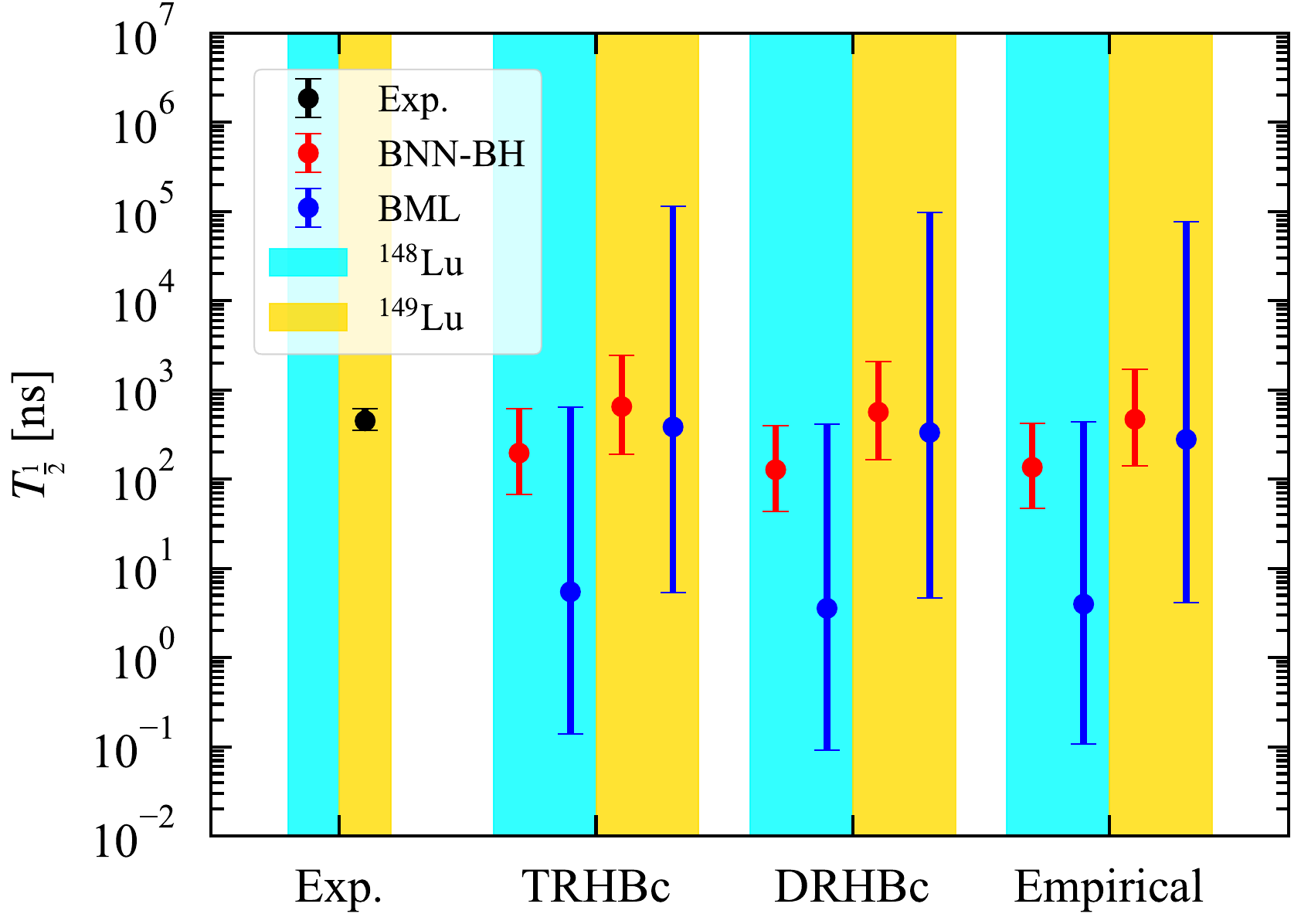}
    \caption{
    Half-lives of proton emitters $^{148}$Lu (cyan region) and $^{149}$Lu (yellow region) in WKB approximation.
    The red solid circles with uncertainties refer to the predictions with $S_{\rm p}$ from the BNN-BH model, and the blue ones refer to those from the BML model~\cite{niu_nuclear_2022}.
    Experimental data are taken from Ref.~\cite{PhysRevLett.128.112501}.
    }\label{pic:lifetime errorbar}
\end{figure}

Adopting the $S_{\rm p}$ of $^{148}$Lu with uncertainty predicted by the BNN-BH and the BML models, the half-lives can be calculated with the potentials from the TRHBc and the DRHBc theories, respectively.
{Table~\ref{tb:result} lists the half-lives of proton emitter $^{148}$Lu in the WKB approximation, along with structural information obtained from the DRHBc and TRHBc calculations. Both theories predict that the valence proton in $^{148}$Lu occupies one orbital dominantly composed of $1h_{11/2}$ components. Although other triaxial minima with different configurations exist near the energy of the ground state, no significant change in proton occupation is expected between $^{149}$Lu and $^{148}$Lu.
Triaxial deformation is found to mainly affect the spectroscopic factor $S_F$ and decay width $\Gamma$, which are consistent with the results for $^{149}$Lu in Ref.~\cite{lu_triaxial_2024}.
It is apparent that the predicted half-lives are highly sensitive to the $Q_{\rm p}$ values, caused by the difference of hundreds of keV between the BNN-BH and the BML models appearing in the exponential power.}  

To clearly exhibit the discrepancy between the predictions and the available experimental data, we plot the half-lives for $^{148}$Lu and $^{149}$Lu isotopes in Fig.~\ref{pic:lifetime errorbar} with the TRHBc+WKB, the DRHBc+WKB models, and an empirical formalism, respectively.
It can be seen that the predicted means of $^{149}$Lu agree well with the experimental data, whether $S_{\rm p}$ came from our BNN-BH or the BML models.
However, the ranges of the half-lives for $^{149}$Lu predicted by WKB approximation with $S_{\rm p}$ from the BNN-BH model span approximately 1 order of magnitude,
while the half-life ranges determined with $S_{\rm p}$ from the BML model span approximately 4 orders of magnitude.
Since our BNN-BH model could predict the $S_{\rm p}$ in a rather stable and precise way with smaller variation compared with the BML model, we have confidence in constraining the half-life of $^{148}$Lu.
The predicted means of the half-lives for $^{148}$Lu by the WKB approximation with $S_{\rm p}$ from the BNN-BH and the BML models differ by approximately two orders of magnitude, and the uncertainties are approximately a factor of three larger using the BML model.



{\it Summary.}---We clarify the shape of $^{148}$Lu nucleus in the ground state as triaxiality and its $Q_{\rm p}$ value as 2.015(89) MeV from our BNN-BH model for the first time.
Taking into account of the ensemble uncertainty and confining the error estimation to the nearby nuclei, our BNN-BH model turns out to effectively constrain $S_{\rm p}$ of $^{148}$Lu in a more stable and precise way than the BML model.
Based on the accuracy of the $Q_{\rm p}$ within 90 keV in our model, the range of the half-life for $^{148}$Lu significantly decreased by 3 orders of magnitude, compared with those by the BML model.
Furthermore, the half-life calculated by the TRHBc + WKB approach is consistent with that from the empirical formalism with the $S_{\rm p}$ predicted by our BNN-BH model.
Since the majority of the predicted values is larger than 100 ns, the half-life of $^{148}$Lu could be possibly measured from the experiment.
In such a way, we cast a new light on the constraint of the half-life for $^{148}$Lu with the help of the BNN-BH model, which might become the next directly measurable proton emitter.
\\ \\
\indent Many thanks to Professors Li-Sheng Geng, Zhong-Ming Niu, Michael Smith and Xiao-Hong Zhou for their helpful discussions.
This work was partly supported by the National Natural Science Foundation of China (Grants Nos.~12575122, 12175010, and 12305125), the National Key Laboratory of Neutron Science and Technology (Grant No.~NST202401016), and the Sichuan Science and Technology Program (Grant No.~2024NSFSC1356).

\bibliography{reference}

\end{document}